\documentclass[sigconf]{acmart}

\setcopyright{none}

\usepackage{subfigure}
\usepackage{enumitem}
\usepackage{booktabs}
\usepackage{balance}
\usepackage{pifont}
\usepackage[linesnumbered,ruled]{algorithm2e}
\usepackage{bm}
\usepackage{breqn}
\usepackage{multirow}

\usepackage{colortbl}

\newcommand{\xhdr}[1]{{\noindent\bfseries #1}.}

\newcommand{\method}{\textit{JODIE}\xspace}
\newcommand{\batching}{\textit{t-Batch}\xspace}

\begin{document}
\title{Predicting Dynamic Embedding Trajectory in\\ Temporal Interaction Networks}

\author{Srijan Kumar}
\affiliation{
\institution{Stanford University, USA and\\ Georgia Institute of Technology, USA}
}
\email{srijan@cs.stanford.edu}
\author{Xikun Zhang}
\affiliation{
\institution{University of Illinois, Urbana-Champaign, USA}
}
\email{xikunz2@illinois.edu}
\author{Jure Leskovec}
\affiliation{
\institution{Stanford University, USA}
}
\email{jure@cs.stanford.edu}

\begin{abstract}
Modeling sequential interactions between users and items/products
is crucial in domains such as e-commerce, social networking, and education. Representation learning presents an attractive opportunity to model the dynamic evolution of users and items, where each user/item can be embedded in a Euclidean space and its evolution can be modeled by an embedding trajectory in this space.
However, existing dynamic embedding methods generate embeddings only when users take actions and do not explicitly model the future trajectory of the user/item in the embedding space.
Here we propose \method, a coupled recurrent neural network model that learns the embedding trajectories of users and items.
\method\ employs two recurrent neural networks to update the embedding of a user and an item at every interaction. Crucially, \method\ also models the future embedding trajectory of a user/item.
To this end, it introduces a novel projection operator that learns to estimate the embedding of the user at any time in the future. These estimated embeddings are then used to predict future user-item interactions. To make the method scalable, we develop a {\em t-Batch} algorithm that
creates time-consistent batches and leads to 9$\times$ faster training. 
We conduct six experiments to validate \method\ on two prediction tasks---future interaction prediction and state change prediction---using four real-world datasets. We show that \method\ outperforms six state-of-the-art algorithms in these tasks by at least 20\% in predicting future interactions and 12\% in state change prediction.

The code and datasets are available on the project website: \\
\url{http://snap.stanford.edu/jodie}.
\end{abstract}

\copyrightyear{2019}
\acmYear{2019}
\setcopyright{acmcopyright}
\acmConference[KDD '19]{The 25th ACM SIGKDD Conference on Knowledge Discovery and Data Mining}{August 4--8, 2019}{Anchorage, AK, USA}
\acmBooktitle{The 25th ACM SIGKDD Conference on Knowledge Discovery and Data Mining (KDD '19), August 4--8, 2019, Anchorage, AK, USA}
\acmPrice{15.00}
\acmDOI{10.1145/3292500.3330895}
\acmISBN{978-1-4503-6201-6/19/08}

\maketitle

\section{Introduction}

\begin{figure}[t]
\centering
\includegraphics[width=\columnwidth]{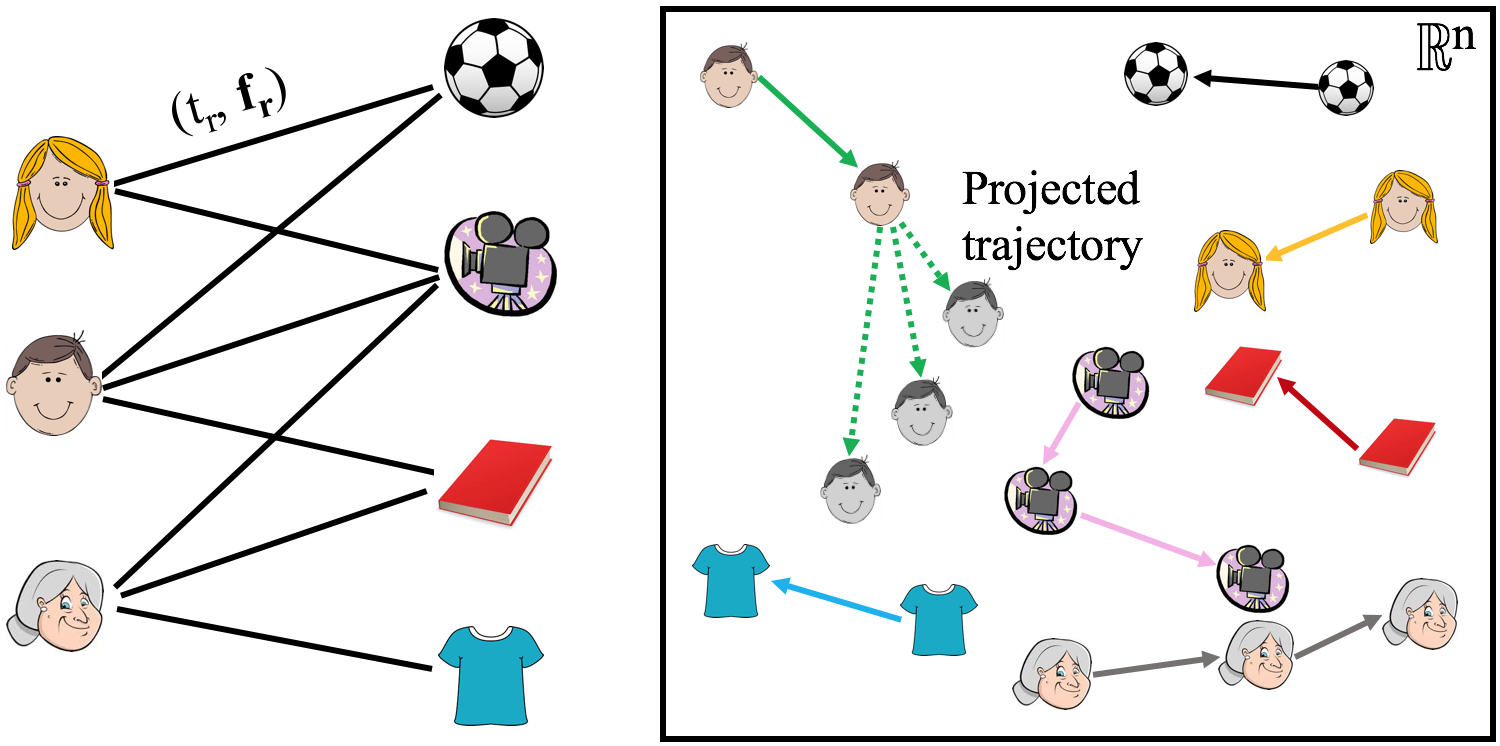}
\vspace{-6mm}
\caption{Left: a temporal interaction network of three users and four items. Each arrow represents an interaction with associated timestamp $t$ and a feature vector $f$. Right: embedding trajectory of the users and items. We predict the future trajectory of users (the dotted line shown for one user) by training an embedding projection operator.
} 
\vspace{-4mm}
\label{fig:toy}
\end{figure}

Users interact sequentially with items in many domains such as e-commerce (e.g., a customer purchasing an item)~\cite{zhang2017deep}, education (a student enrolling in a MOOC course)~\cite{liyanagunawardena2013moocs}, and social and collaborative platforms (a user posting in a group in Reddit)~\cite{iba2010analyzing,kumar2018community}. 
The same user may interact with different items over a period of time and these interactions change over time~\cite{DBLP:journals/debu/HamiltonYL17,DBLP:conf/recsys/PalovicsBKKF14,zhang2017deep,agrawal2014big,DBLP:conf/asunam/ArnouxTL17,raghavan2014modeling,DBLP:journals/corr/abs-1711-10967}.
These interactions create a \textit{network of temporal interactions between users and items}.
Figure~\ref{fig:toy} (left) shows an example network between users and items, with each interaction 
marked with a time stamp $t_r$ and a feature vector $\bm{f_r}$ (such as the review text or the purchase amount).
Accurate real-time recommendation of items and predicting change in the state of users are fundamental problems in these domains~\cite{DBLP:conf/wsdm/QiuDMLWT18,DBLP:conf/asunam/ArnouxTL17,DBLP:conf/sdm/LiDLLGZ14,DBLP:journals/corr/abs-1804-01465,DBLP:conf/cosn/SedhainSXKTC13,walker2015complex,DBLP:conf/icwsm/Junuthula0D18}.
For instance, predicting when a student is likely to drop out of a MOOC course 	is important to develop early intervention measures~\cite{kloft2014predicting,yang2013turn} and predicting when a user is likely to turn malicious on social platforms, like Reddit and Wikipedia, ensures platform integrity~\cite{kumar2018rev2,kumar2015vews,cheng2017anyone}.

Representation learning, or learning low-dimensional embeddings of entities, is a powerful approach to represent the evolution of users' and items' properties~\cite{DBLP:journals/kbs/GoyalF18,zhang2017deep,dai2016deep,DBLP:conf/nips/FarajtabarWGLZS15,beutel2018latent,zhou2018dynamic}. 
However, the recent methods that generate dynamic embeddings suffer from four fundamental challenges.
\textbf{First,} a majority of the existing methods generate an embedding for a user only when she takes an action~\cite{beutel2018latent,zhang2017deep,dai2016deep,zhou2018dynamic,you2019hierarchical}.
However, consider a user who makes a purchase today and its embedding is updated. The embedding will remain the same if it returns to the platform on the next day, a week later, or even a month later. As a result, the same predictions and recommendations will be made to her regardless of when she returns. However, a user's intent changes over time~\cite{cheng2017predicting} and thus her embedding needs to be updated (projected) to the query time. The challenge here is how to accurately predict the embedding trajectories of users/items as time progresses.
\textbf{Second}, entities have both stationary properties that do not change over time and time-evolving properties. Some existing methods~\cite{zhang2017deep,dai2016deep,wang2016coevolutionary} consider only one of the two when generating embeddings. However, it is essential to consider both in a unified framework to leverage information at both scales.
\textbf{Third}, many existing methods predict user-item interactions by scoring all items for each user~\cite{zhang2017deep,dai2016deep,beutel2018latent,zhou2018dynamic}. This has linear time complexity and is not practical in scenarios with millions of items. Instead, methods are required that can recommend items in near-constant time.
\textbf{Fourth}, most models are trained by sequentially processing the interactions one at a time, so that the temporal dependencies between the interactions are maintained~\cite{zhang2017deep,dai2016deep,wang2016coevolutionary}. This prevents such models from scaling to datasets with millions of interactions. New methods are needed that can be trained with batches of data to generate embedding trajectories.

\textbf{Present work.}
Here we present \method\ which learns to generate embedding trajectories of all users and items from temporal interactions\footnote{\method\ stands for \underline{Jo}int \underline{D}ynamic User-\underline{I}tem \underline{E}mbeddings.}.
The embedding trajectories of the example network are shown in Figure~\ref{fig:toy} (right).
The embeddings of the user and item are updated when a user takes an action and a projection operator predicts the future embedding trajectory of the user.

\textbf{Present work: \method.}
Each user and item has two embeddings: a static embedding and a dynamic embedding.
The static embedding represents the entity's long-term stationary property, while the dynamic embedding represents time-varying property and is learned using the \method\ algorithm.
Both embeddings are used to generate the trajectory.
This enables \method\ to make predictions from both the stationary and time-varying properties of the user.

The \method\ model consists of two major components: an update operation and a projection operation. 

The \textit{update operation} of \method\ has two Recurrent Neural Networks (RNNs) to generate user and item embeddings. 
Crucially, the two RNNs are coupled to explicitly incorporate the interdependency between users and items. 
After each interaction, the user RNN updates the user embedding by using the embedding of the interacting item. Similarly, the item RNN uses the user embedding to update the item embedding. 
The model also has the ability to incorporate feature vectors from the interaction, for example, the text of a Reddit post.
It should be noted that \method\ is easily extendable to multiple types of entities by training one RNN for each entity type.
In the current work, we show how to apply \method\ to the case of bipartite interactions between users and items.

A major innovation of \method\ is that it also uses a \textit{projection operation} that predicts the future embedding trajectory of the users. 
Intuitively, the embedding of a user will change slightly after a short time elapses since her previous interaction (with any item), while the embedding can change significantly after a long time elapses.
As a result, \method\ trains a temporal attention layer to project the embedding of users after some time $\Delta$ elapses since its previous interaction.
The projected user embedding is then used to predict the item that the user is most likely to interact with.

To predict the item that a user will interact with, an important design decision is to output the embedding of an item, instead of an interaction probability.
Current methods generate a probability score of interaction between a user and an item, which takes linear time to find the most likely item because probability scores for all items have to be generated first.
Instead, by directly generating the item embedding, we can recommend the item that is closest to the predicted item in the embedding space. This can be done efficiently in constant time using the locality sensitive hashing (LSH) techniques~\cite{leskovec2014mining}.

\textbf{Present work: \batching.}
Most existing models learn embeddings from a sequence of interactions by processing one interaction after the other, in increasing order of time to maintain the temporal dependency among the interactions~\cite{dai2016deep, zhang2017learning,wang2016coevolutionary}.
This makes such algorithms unscalable to real datasets with millions of interactions.
Therefore, we create a batching algorithm, called \batching, to train \method\ by creating training batches of independent interactions such that the interactions in each batch can be processed in parallel. 
To do so, we iteratively select independent edge sets from the interaction network.
In every batch, each user and item appears at most once and the temporally-sorted interactions of each user (and item) appear in monotonically increasing batches.
Experimentally, we show that \batching\ makes \method\ 9.2$\times$ faster than its most similar dynamic embedding baselines. 

\textbf{Present work: Experiments.}
We conduct six experiments to evaluate the performance of \method\ on two tasks: predicting the next interaction of a user and predicting the change in state of users (when a user will be banned from social platforms and when a student will drop out from a MOOC course). 
We use four datasets from Reddit, Wikipedia, LastFM, and a MOOC course activity for our experiments.
We compare \method\ with six state-of-the-art algorithms from three categories: deep recurrent recommender algorithms~\cite{zhu2017next,beutel2018latent,wu2017recurrent}, temporal node embedding algorithm~\cite{nguyen2018continuous}, and dynamic co-evolution models~\cite{dai2016deep}.
\method\ improves over the baseline algorithms on the interaction prediction task by at least 20\% in terms of mean reciprocal rank and 12\% in AUC on average for predicting user state change. 
We further show that \method\ is robust to the percentage of training data and the size of the embeddings. 

Overall, in this paper, we make the following contributions:\\
$\bullet$ \textbf{Embedding algorithm:} We propose a coupled recurrent neural network model called \method to learn embedding trajectories of users and items. Crucially, \method also learns a projection operator to predict the embedding trajectory of users and predicts future interactions in constant time.\\ 
$\bullet$ \textbf{Batching algorithm:} We propose the \batching\ algorithm to create independent but temporally consistent training data batches that help to train \method\ 9.2$\times$ faster than the closest baseline.\\
$\bullet$ \textbf{Effectiveness:} \method\ outperforms six state-of-the-art algorithms in predicting future interactions and user state change predictions, by performing at least 20\% better in predicting future interactions and 12\% better on average in predicting user state change. 

The code and datasets are available on the project website: \\\texttt{https://snap.stanford.edu/jodie}.

\section{Related Work}
\label{sec:related}

Here we discuss the research closest to our problem setting spanning three broad areas. 
Table~\ref{tab:related} compares their differences.

{
\footnotesize
\begin{table}[t]
\center
\caption{\label{tab:related} Table comparing the desired properties of the existing algorithms and our proposed \method\ algorithm. \method\ satisfies all the desirable properties.}
\vspace{-3mm}
\begin{tabular}{c|c|c|c|c|c|c||c}
\hline
& \multicolumn{3}{c|}{Deep} & \multicolumn{2}{c|}{Temporal} & \multirow{4}{*}{\rotatebox[origin=c]{90}{Co-evolution models~\cite{dai2016deep}\quad}} & Proposed \\
& \multicolumn{3}{c|}{recurrent} & \multicolumn{2}{c|}{network} & & model \\
& \multicolumn{3}{c|}{models} & \multicolumn{2}{c|}{embedding} & & \\\cline{2-6}\cline{8-8}
Property & \rotatebox[origin=c]{90}{LSTM, Time-LSTM~\cite{zhu2017next}} & \rotatebox[origin=c]{90}{RRN~\cite{wu2017recurrent}} & \rotatebox[origin=c]{90}{LatentCross~\cite{beutel2018latent}} & \rotatebox[origin=c]{90}{CTDNE~\cite{nguyen2018continuous}} & \rotatebox[origin=c]{90}{IGE~\cite{zhang2017learning}} & & \rotatebox[origin=c]{90}{\method} \\\hline
Predict embedding trajectory & & & & & & & \ding{52} \\
Predict future item embedding & & & & & & & \ding{52} \\
Train using batches of data & \ding{52}  &  \ding{52} & \ding{52}  &\ding{52} & & & \ding{52} \\\hline
\end{tabular}
\vspace{-3mm}
\end{table}
}

\textbf{Deep recurrent  recommender models.}
Several recent models employ recurrent neural networks (RNNs) and variants (LSTMs and GRUs) to build recommender systems.
RRN~\cite{wu2017recurrent} uses RNNs to generate dynamic user and item embeddings from rating networks.
Recent methods, such as Time-LSTM~\cite{zhu2017next} and LatentCross~\cite{beutel2018latent} learn how to incorporate features into the embeddings.
However, most of these methods suffer from two major shortcomings. First, they take the one-hot vector of the item as input to update the user embedding.
This only incorporates the item id and ignores the item's current state. 
The second shortcoming is that some models, such as Time-LSTM and LatentCross, generate dynamic embeddings only for users and not for items. 

\method\ overcomes these shortcomings by learning embeddings for both users and items using mutually-recursive RNNs.
In doing so, \method\ outperforms these methods by at least 20\% in predicting the next interaction and 12\% on average in predicting user state change, while having comparable running time as these methods.

\textbf{Dynamic co-evolution models.}
Methods that jointly learn representations of users and items have recently been developed using point-process modeling~\cite{wang2016coevolutionary,trivedi2017know} and RNN-based modeling~\cite{dai2016deep}.
The basic idea behind these models is similar to \method\ ---user and item embeddings influence each other whenever they interact.
However, the major difference between \method\ and these models are that \method\ trains a project operation to forecast the user embedding at any time, outputs item embeddings instead of interaction probability, and trains the model using batching.
As a result, we observe that \method\ outperforms DeepCoevolve by at least 44.8\% in predicting the next interaction and 14\% in predicting state change.
In addition, most of these existing models are not scalable because they process interactions in a sequential order to maintain temporal dependency.
\method\ overcomes this limitation by creating efficient training data batches which makes \method\ 9$\times$ faster than these baselines.

\textbf{Temporal network embedding models.}
Several models have recently been developed that generate embeddings for the nodes (users and items) in temporal networks.
CTDNE~\cite{nguyen2018continuous} is a state-of-the-art algorithm that generates embeddings using temporally-increasing random walks, but it generates one final static embedding of the nodes.
Similarly, IGE~\cite{zhang2017learning} generates one final embedding of users and items from interaction graphs. 
Therefore, both these methods (CTDNE and IGE) need to be re-run for every new edge to create dynamic embeddings.
Another recent algorithm, DynamicTriad~\cite{zhou2018dynamic} learns dynamic embeddings but does not work on bipartite interaction networks as it requires the presence of triads.
Other recent algorithms such as DDNE~\cite{DBLP:journals/access/LiZYZY18}, DANE~\cite{DBLP:conf/cikm/LiDHTCL17}, DynGem~\cite{goyal2018dyngem}, Zhu et al.~\cite{zhu2016scalable}, and Rahman et al.~\cite{DBLP:journals/corr/abs-1804-05755} learn embeddings from a sequence of graph snapshots, which is not applicable to our setting of continuous interaction data.
Recent models such as NP-GLM model~\cite{DBLP:journals/corr/abs-1710-00818}, DGNN~\cite{DBLP:journals/corr/abs-1810-10627}, and DyRep~\cite{trivedi2018representation} learn embeddings from persistent links between nodes, which do not exist in interaction networks as the edges represent instantaneous interactions.

Our proposed model, \method\, overcomes these shortcomings by generating and predicting the trajectories of users and items. In doing so, \method\ performs 4.4$\times$ better than CTDNE in predicting the next interaction, while having comparable running time.

\section{JODIE: Joint Dynamic User-Item Embedding Model}
\label{sec:method}

\begin{figure}[t]
\centering
\includegraphics[width=0.8\columnwidth]{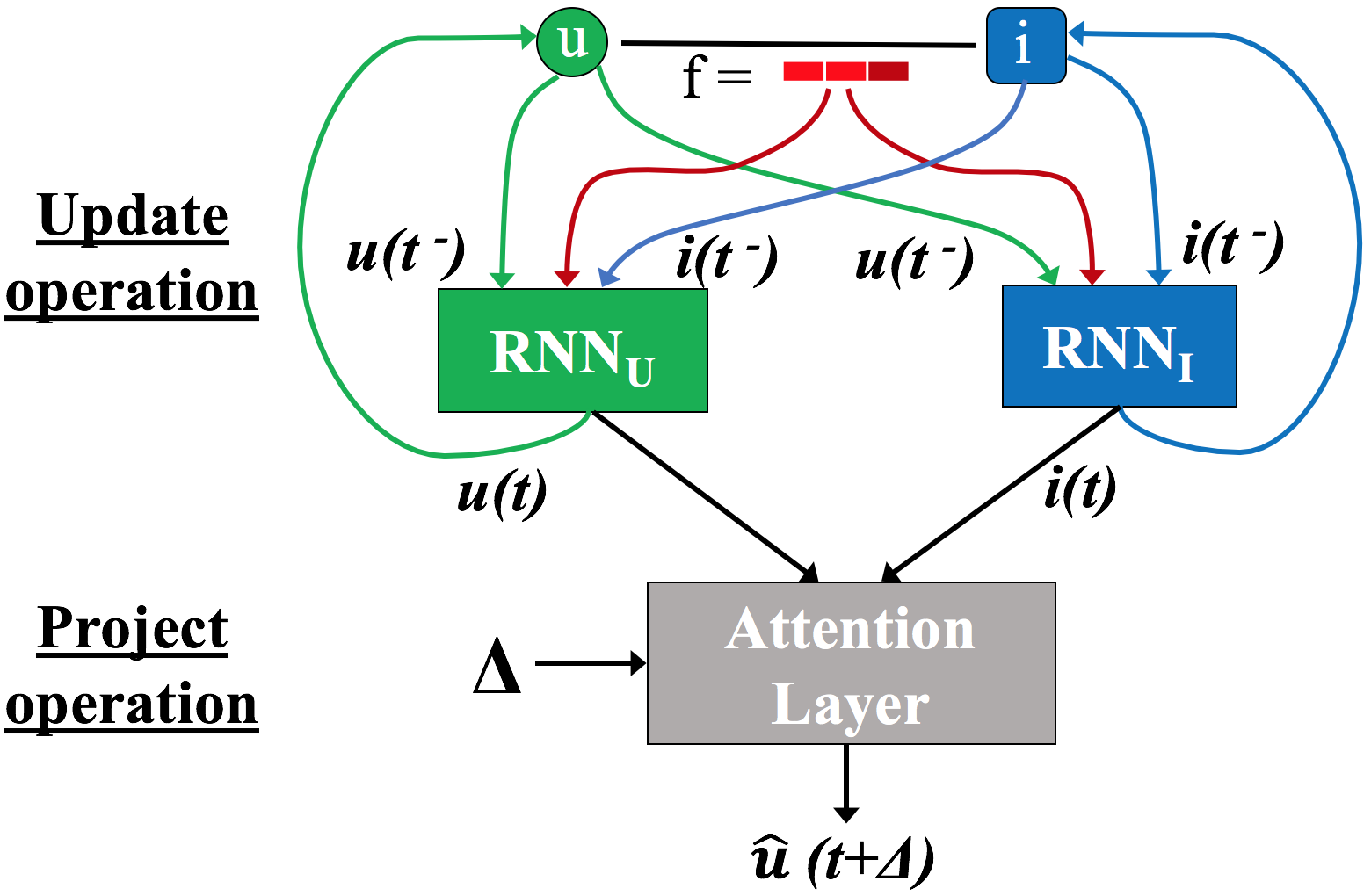}
\vspace{-2mm}
\caption{\textbf{The \method\ model:} After an interaction $(u, i, t, f)$ between user $u$ and item $i$, the dynamic embeddings of $u$ and $i$ are updated in the \textit{update operation} with $RNN_U$ and $RNN_I$, respectively.
The \textit{projection operation} predicts the user embedding at a future time $t+\Delta$. 
\label{fig:model}}
\vspace{-2mm}
\end{figure}

In this section, we propose \method, a method to learn embedding trajectories of users $\bm{u(t)} \in  \mathbb{R}^n  \text{ } \forall u \in \mathcal{U}$ and items $\bm{i(t)} \in \mathbb{R}^n  \text{ } \forall i \in \mathcal{I}, \forall t \in [0, T]$ from an ordered sequence of temporal user-item interactions $S_r = (u_r, i_r, t_r, \bm{f_r})$.
An interaction $S_r$ happens between a user $u_r \in \mathcal{U}$ and an item $i_r \in \mathcal{I}$ at time $t_r \in  \mathbb{R}^+, 0 < t_1 \leq t_2 \ldots \leq T$.
Each interaction has an associated feature vector $\bm{f_r}$ (e.g., a vector representing the text of a post).
Table~\ref{tab:symbols} lists the symbols used.
For ease of notation, we will drop the subscript $r$ in the rest of the section.

\begin{figure}[t!]
\centering
\includegraphics[width=0.8\columnwidth]{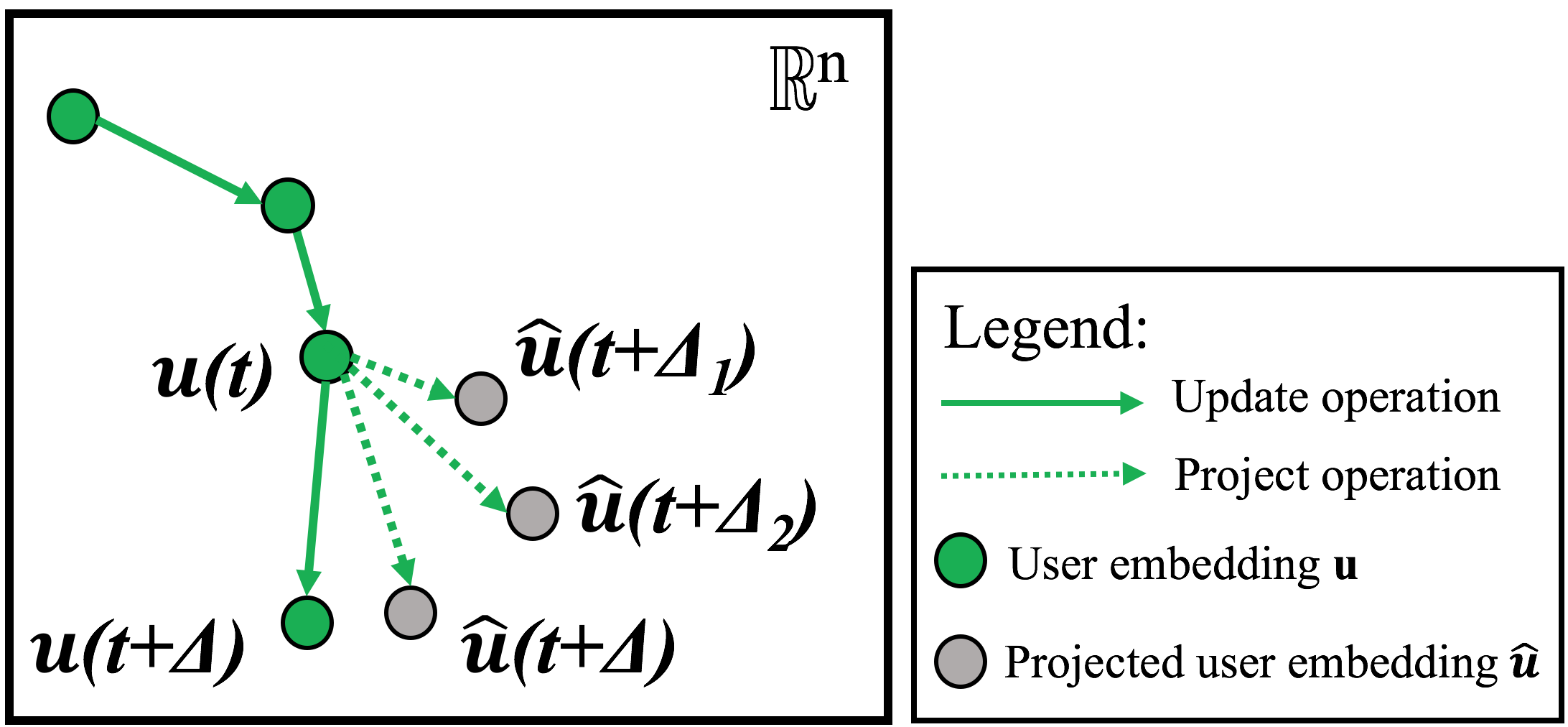}
\vspace{-3mm}
\caption{This figure shows the key idea behind \textit{projection operation}. The predicted embedding of user $u$ is shown for different elapsed time $\Delta_1 < \Delta_2 < \Delta$. The predicted embedding drifts farther as more time elapses. When the next interaction is observed, the embedding is updated again.}
\vspace{-3mm}
\label{fig:projection}
\end{figure}

Our proposed model, called \method, learns an embedding trajectory for users and items and is reminiscent of the popular Kalman Filtering algorithm~\cite{julier1997new}.\footnote{Kalman filtering is used to accurately measure the state of a system using a combination of system observations and state estimates given by the laws of the system.}
\method\ uses the interactions to update the state of the interacting users and items via a trained \textit{update operation}.
\method\ trains a \textit{projection operation} that uses the previous observed state and the elapsed time to predict the future embedding of the user.
When the user's and item's next interactions are observed, their embeddings are updated again.
We illustrate the model in Figure~\ref{fig:model} and the projection operation in Figure~\ref{fig:projection}.

\textbf{Static and Dynamic Embeddings.}
Each user and item is assigned two embeddings: a static and a dynamic embedding.
We use both embeddings to encode both the long-term stationary properties of the entities and their dynamic properties.

Static embeddings, $\bm{\overline{u}} \in \mathbb{R}^d \text{ } \forall u \in \mathcal{U}$ and $\bm{\overline{i}} \in \mathbb{R}^d \text{ } \forall i \in \mathcal{I}$, do not change over time.
These are used to express stationary properties such as the long-term interest of users.
We use one-hot vectors as static embeddings of all users and items, as advised in Time-LSTM~\cite{zhu2017next} and TimeAware-LSTM~\cite{baytas2017patient}.
Using node2vec~\cite{grover2016node2vec} gave empirically similar results, so we use one-hot vectors.

On the other hand, each user $u$ and item $i$ is assigned a dynamic embedding represented as $\bm{u(t)} \in \mathbb{R}^n$ and $\bm{i(t)} \in \mathbb{R}^n$ at time $t$, respectively.
These embeddings change over time to model their time-varying behavior and properties.
The sequence of dynamic embeddings of a user/item is referred to its \textit{trajectory}.

Next, we describe the update and projection operations. Then, we will describe how we predict the future interaction item embeddings and how we train the model.

\vspace{-2mm}
\subsection{\textbf{Embedding update operation}}
\label{sec:update}
In the update operation, the interaction $S = (u, i, t, \bm{f})$ between a user $u$ and item $i$ at time $t$ is used to generate their dynamic embeddings $\bm{u(t)}$ and $\bm{i(t)}$. Fig.~\ref{fig:model} illustrates the update operations.

Our model uses two recurrent neural networks for updates---$RNN_U$ is shared across all users to update user embeddings, and $RNN_I$ is shared among all items to update item embeddings.
The hidden states of the user RNN and the item RNN represent the user and item embeddings, respectively.

The two RNNs are mutually-recursive.
When user $u$ interacts with item $i$, $RNN_U$ updates the embedding $\bm{u(t)}$ by using the embedding $\bm{i(t^-)}$ of item $i$ right before time $t$ as an input. $\bm{i(t^-)}$ is the same as item $i$'s embedding after its previous interaction with any user.
Notice that this design decision is in stark contrast with the popular use of items' one-hot vectors to update user embeddings~\cite{beutel2018latent,wu2017recurrent,zhu2017next}, which has the following two disadvantages: (a) one-hot vector only contains the information about the item's id and not the item's current state, and (b) the dimension of the one-hot vector becomes very large when real datasets have millions of items, making the model challenging to train and scale.
Instead, we use the dynamic embedding of an item as it reflects the item's current state leading to more meaningful dynamic user embeddings and easier training.
For the same reason, $RNN_I$ updates the dynamic embedding $\bm{i(t)}$ of item $i$ by using the dynamic user embedding $\bm{u(t^-)}$ (which is $u$'s embedding right before time $t$).
This results in mutually recursive dependency between the embeddings. More formally,
$$ \bm{u(t)} =  \sigma(W_1^{u}\bm{u(t^-)} + W_2^{u}\bm{i(t^-)} + W_3^{u}\bm{f} + W_4^u \Delta_u) $$
$$ \bm{i(t)} =   \sigma(W_1^{i}\bm{i(t^-)} + W_2^{i}\bm{u(t^-)} + W_3^{i}\bm{f} + W_4^i \Delta_i) $$
where $\Delta_u$ denotes the time since $u$'s previous interaction (with any item) and $\Delta_i$ is the time since item $i$'s previous interaction (with any user). $\bm{f}$ is the interaction feature vector. The matrices $W_1^u, \ldots W_4^u$ are the parameters of $RNN_U$ and matrices $W_1^i, \ldots W_4^i$ are the parameters of $RNN_I$. $\sigma$ is a sigmoid function to introduce non-linearity. The matrices are trained to predict the embedding of the item at $u$'s next interaction as explained later in Section~\ref{sec:prediction}.

Variants of RNNs, such as LSTM, GRU, and T-LSTM~\cite{zhu2017next}, gave experimentally similar and sometimes worse performance, so we use RNNs in our model to reduce the number of trainable parameters.

\begin{table}
\caption{\label{tab:symbols} Table of symbols used in this paper.}
\vspace{-2mm}
\begin{tabular}{|c|l|}
\hline
Symbol & Meaning \\\hline
$\bm{u(t)}$ and $\bm{i(t)}$ & Dynamic embedding of user $u$ and item $i$ at time $t$ \\
$\bm{u(t^-)}$ and $\bm{i(t^-)}$ & Dynamic embedding of user $u$ and item $i$ before time $t$ \\
$\bm{\overline{u}}$ and $\bm{\overline{i}}$ & Static embedding of user $u$ and item $i$ \\
$\bm{\widehat{u}(t)}$ & Projected embedding of user $u$ at time $t$ \\
$\bm{\widetilde{j}(t)}$ & Predicted item $j$ embedding\\\hline
\end{tabular}
\end{table}

\vspace{-2mm}
\subsection{\textbf{Embedding projection operation}}
\label{sec:project}
Here we explain one of the major contributions of our algorithm, the embedding projection operator, which predicts the future embedding trajectory of the user. This is done by projecting the embedding of the user at a future time.
The projected embedding can then be used for downstream tasks, such as predicting items the user will interact with 
at a given query/prediction time in the future.

Figure~\ref{fig:projection} visualizes the main idea of projecting a user's embedding trajectory. 
The operation projects the embedding of a user after some time has elapsed since its last interaction at time $t$.
To give an example, a short duration $\Delta_1$ after time $t$, the user $u$'s projected embedding $\bm{\widehat{u}(t+\Delta_1)}$ is close to its previously observed embedding $\bm{u(t)}$. As more time $\Delta > \Delta_2 > \Delta_1$ elapses, the projected embeddings drift farther to $\bm{\widehat{u}(t+\Delta_2)}$ and $\bm{\widehat{u}(t+\Delta)}$. When the next interaction is observed at time $t + \Delta$, the user's embedding is updated to $\bm{u(t + \Delta)}$ using the update operation.

Two inputs are required for the projection operation: $u$'s embedding at time $t$ and the elapsed time $\Delta$.
We follow the method suggested in LatentCross~\cite{beutel2018latent} to incorporate time into the projected embedding via Hadamard product. We do not simply concatenate the embedding and the time and pass them through a linear layer as prior research has shown that neural networks are inefficient in modeling the interactions between concatenated inputs. Instead, we create a temporal attention vector as described below.

We first convert $\Delta$ to a time-context vector $\bm{w} \in \mathbb{R}^n$ using a linear layer (represented by vector $W_p$): $ \bm{w} = W_p \Delta$.
We initialize $W_p$ by a 0-mean Gaussian.
The projected embedding is then obtained as an element-wise product of the time-context vector with the previous embedding as follows:
$$\bm{\widehat{u}(t + \Delta)} = (1 + \bm{w}) * \bm{u(t)}$$
The vector $1 + \bm{w}$ acts as a temporal attention vector to scale the past user embedding. When $\Delta = 0$, then $\bm{w} = 0$ and the projected embedding is the same as the input embedding vector. The larger the value of $\Delta$, the more the projected embedding vector differs from the input embedding vector and the projected embedding vector drifts over time.

We find that a linear layer works the best to project the embedding as it is equivalent to a linear transformation in the embedding space. Adding non-linearity to the transformation makes the projection operation non-linear, which we find experimentally to reduce the prediction performance.
Thus, we use the linear transformation as described above.

Next, we describe how we train the model to efficiently project user embeddings such that they are useful in predicting the next item with which the user will interact.

\vspace{-2mm}
\subsection{\textbf{Training to predict next item embedding}}
\label{sec:prediction}
Let $u$ interact with item $i$ at time $t$ and then with item $j$ at time $t + \Delta$. Right before $t + \Delta$, can we predict which item $u$ will interact with? We use this task to train the update and projection operations in \method.
We train \method\ to make this prediction using $u$'s projected embedding $\bm{\widehat{u}(t + \Delta)}$.

A crucial design decision here is that \method\ directly outputs an item embedding vector, $\bm{\tilde{j}(t + \Delta)}$, instead of an interaction probability between $u$ and item $j$.
This has the advantage of reducing the computation at inference time from linear (in the number of items) to near-constant.
Most existing methods~\cite{dai2016deep,wu2017recurrent,beutel2018latent,DBLP:conf/kdd/DuDTUGS16} that output an interaction probability need to do the expensive neural-network forward pass $|\mathcal{I}|$ times (once for each of item $\in \mathcal{I}$) to find the item with the highest probability score.
In contrast, \method\ only needs to do forward-pass of the prediction layer once and output a predicted item embedding. Then the item with the closest embedding can be returned in near-constant time by using Locality Sensitive Hashing (LSH) techniques~\cite{leskovec2014mining}. To maintain the LSH data structure, we update it whenever an item's embedding is updated.

Thus, we train \method\ to minimize the $L_2$ difference between the predicted item embedding $\bm{\tilde{j}(t + \Delta)}$ and the real item embedding $[\bm{\overline{j}}, \bm{j(t+\Delta^-)}]$ as follows: $|| \bm{\tilde{j}(t + \Delta)} - \bm{[\overline{j}}, \bm{j(t + \Delta^-)]} ||_2$. Here, $\bm{[x,y]}$ represents the concatenation of vectors $\bm{x}$ and $\bm{y}$, and the superscript `-' indicates the embedding immediately before the time.

We make this prediction using the projected user embedding $\bm{\widehat{u}(t + \Delta)}$ and the embedding $\bm{i(t+\Delta^-)}$ of item $i$ (the item from $u$'s previous interaction) immediately before time $t + \Delta$.
The reason we include $\bm{i(t+\Delta^-)}$ is two-fold: (a) $i$ may interact with other users between time $t$ and $t + \Delta$, and thus the embedding contains more recent information, and (b) users often interact with the same item consecutively (i.e., $i = j$) and including the item embedding helps to ease the prediction.
We use both the static and dynamic embeddings to predict the static and dynamic embedding of the predicted item $j$.
The prediction is made using a fully connected linear layer as follows:
$$ \bm{\tilde{j}(t + \Delta)} = W_1\bm{\widehat{u}(t + \Delta)} + W_2 \bm{\overline{u}} + W_3 \bm{i(t + \Delta^-)} + W_4 \bm{\overline{i}} + B $$
where $W_1, \ldots W_4$ and the bias vector $B$ make the linear layer.

\textbf{Training the model.}
\method\ is trained to minimize the $L_2$ distance between the predicted item embedding and the ground truth item's embedding at every interaction.
We calculate the total loss as follows:
\begin{dmath}
$$
Loss =  \sum\limits_{(u,i,t,\bm{f}) \in S} || \bm{\tilde{j}(t)} - \bm{[\overline{i}}, \bm{i(t^-)]} ||_2 $$
$$ \text{ } + \text{ } \lambda_U ||\bm{u(t)} - \bm{u(t^-)} ||_2 \text{ } + \text{ }  \lambda_I || \bm{i(t)} - \bm{i(t^-)} ||_2 $$
\end{dmath}

The first loss term minimizes the predicted embedding error. The last two terms are added to regularize the loss and prevent the consecutive dynamic embeddings of a user and item to vary too much, respectively. 
$\lambda_U$ and $\lambda_I$ are scaling parameters to ensure the losses are in the same range. 
It is noteworthy that we do not use negative sampling during training as \method\ directly outputs the embedding of the predicted item. 

\xhdr{Extending the loss for categorical prediction}
In certain prediction tasks, such as user state change prediction, additional training labels may be present for supervision.
The user state change labels are binary (categorical).
In those cases, we can train another prediction function $\Theta: \mathbb{R}^{n+d} \rightarrow \mathcal{C}$ to predict the label using the embedding of the user after an interaction.
We calculate the cross-entropy loss for categorical labels and add the loss to the above loss function with another scaling parameter.
We explicitly do not just train to minimize only the cross-entropy loss to prevent overfitting.

\vspace{-2mm}
\subsection{t-Batch: Training data batching}
\label{sec:batching}

Here we explain the batching algorithm we propose to parallelize the training of \method. It is important to maintain temporal dependencies between interactions during training, such that interaction $S_r$ is processed before $S_k$ $\forall r < k$.

Existing methods that use a single RNN, such as T-LSTM~\cite{zhu2017next} and RRN~\cite{beutel2018latent}, split users into different batches and process them in parallel. This is possible because these approaches use one-hot vector encodings of items as inputs and can thus be trained using the standard Back Propagation Through Time (BPTT) mechanism.

However, in \method, the mutually-recursive RNNs enable us to incorporate the item's embedding to update the user embedding and vice-versa. This creates interdependencies between two users that interacted with the same item and this prevents us from simply splitting users into separate batches and processing them in parallel.

Most existing methods that also use two mutually-recursive RNNs~\cite{dai2016deep,zhang2017learning} naively process all the interactions one at a time in sequential order. However, this is not scalable to a large number of interactions as the training process is very slow. Therefore, we train \method\ using a training data batching algorithm that we call \batching. This leads to an order of magnitude of speed-up in \method\ compared to most existing training approaches.

Creating the training batches is challenging because it has two requirements: (1) all interactions in each batch should be processed in parallel, and (2) processing the batches in increasing order of their index should maintain the temporal ordering of the interactions and thus, it should generate the same embedding as without any batching.

To overcome these challenges, \batching\ creates each batch by selecting independent edge sets of the interaction network, i.e., two interactions in the same batch do not share any common user or item.
\method\ works iteratively in two steps: the select step and the reduce step.
In the \textit{select step}, a new batch is created by selecting the maximal edge set such that each edge $(u,i)$ is the lowest time-stamped edge incident on both $u$ and $i$. This trivially makes the batch an independent edge set.
In the \textit{reduce step}, the selected edges are removed from the network.
\method\ iterates the two steps till no edges remain in the graph.
Thus, each batch is parallelizable and processing batches in order maintains the sequential dependencies.

In practice, we implement \batching\ as a sequential algorithm as follows.
The algorithm assigns each interaction $S_r$ to a batch $B_k$, where $k \in [1, |\mathcal{I}|]$.
We initialize $|\mathcal{I}|$ empty batches (in the worst case scenario that each batch only has one interaction).
We iterate through the temporally-sorted sequence of interactions $S_1 \ldots S_{|\mathcal{I}|}$ and add each interaction to a batch $B_k$.
Let $maxBatch(e, r)$ be the batch with the largest index that has an interaction involving an entity $e$ till interaction $S_r$.
Then, the interaction $S_{r+1}$ (say, between user $u$ and item $i$) is assigned to the batch with index $= max( 1 + maxBatch(u, r), 1 + maxBatch(i, r))$.
The complexity of creating the batches is $\mathcal{O}(|S|)$, i.e., linear in the number of interactions, as each interaction is used once.

It is trivial to verify that \batching\ algorithm satisfies the two requirements.
\batching\ ensures that each user and item appears at most once in every batch and thus, each batch can be parallelized.
In addition, the $r^{th}$ and $r+1^{st}$ interactions of every user and every item are assigned to batches $B_k$ and $B_l$, respectively, such that $k < l$.
So, \method\ can process the batches in increasing order of their indices to ensure that the temporal ordering of the transactions is respected.

We do not predetermine the number and size of the batches because it depends on the interactions in the dataset. The number of batches can range between 1 and $|\mathcal{I}|$. Let us illustrate these two extreme cases. When all interactions have unique users and items, then only one batch is created that has all the interactions. On the other extreme, if all interactions are associated to the same user or the same item, then $|\mathcal{I}|$ batches are created. Therefore, we initialize $|\mathcal{I}|$ batches and discard all trailing empty batches after assignment.

\vspace{-2mm}
\subsection{\textbf{Differences between \method and DeepCoevolve}}
DeepCoevolve is the closest state-of-the-art algorithm to \method because it also trains two mutually-recursive RNNs to generate embedding trajectories.
However, the key differences between \method\ and DeepCoevolve are the following: (i) \method\ uses a novel project function to predict the future trajectory of users. Instead, DeepCoevolve maintains the same embedding of a user between two of its consecutive interactions. Predicting the trajectory enables \method\ to make more effective predictions. (ii) \method\ predicts the embedding of the next item that a user will interact with. In contrast, DeepCoevolve predicts the probability of interaction between a user and an item. During inference time, DeepCoevolve requires $|\mathcal{I}|$ forward passes through the inference layer (for $|\mathcal{I}|$ items) to recommend the item with the highest score. On the other hand, \method\ takes near-constant time. (iii) \method\ is trained with batches of interaction data, as opposed to individual interactions.

As a result, as we will see in the experiments section, \method\ significantly outperforms DeepCoevolve both in terms of performance and training time. \method\ is 9.2$\times$ faster, 45\% better in predicting future interactions, and 13.9\% better in predicting user state change on average.

\section{Experiments}
\label{sec:experiments}

In this section, we experimentally validate the effectiveness of \method\ on two tasks: future interaction prediction and user state change prediction.
We conduct experiments on three datasets each and compare with six strong baselines to show the following:
\begin{enumerate}
\item \method\ outperforms the baselines by at least 20\% in terms of mean reciprocal rank in predicting the next item and 12\% on average in predicting user state change.
\item  We show that \method\ is 9.2$\times$ faster than DeepCoevolve and comparable to other baselines.
\item  \method\ is robust in performance to the availability of training data and the dimension of the embedding.
\item Finally, in a case study on the MOOC dataset, we show that \method\ can predict student drop-out five interactions in advance.
\end{enumerate}

We first explain the experimental setting and the baseline methods and then describe the experimental results.

\textbf{Experimental setting.}
We train all models by splitting the data by time to simulate the real situation.
Thus, we train all models on the first $\tau\%$ interactions, validate on the next $\tau_v\%$, and test on the last remaining interactions.

For a fair comparison, we use 128 dimensions as the dimensionality of the dynamic embedding for all algorithms and one-hot vectors for static embeddings.
All algorithms are run for 50 epochs, and all reported numbers for all models are for the test data corresponding to the best performing validation set.

{
\begin{table*}
\small
\caption{\label{tab:interaction} \textbf{Future interaction prediction experiment:} Table comparing the performance of \method\ with state-of-the-art algorithms, in terms of mean reciprocal rank (MRR) and recall@10.  The {\color{blue!75}best algorithm} in each column is colored {\color{blue!75}blue} and {\color{blue!20}second best is light blue}. The last two columns show the minimum percentage improvement of \method\ over the method, across over all datasets. We see that \method\ outperforms all baselines by at least 20\% in MRR and 14\% in recall@10.}
\vspace{-2mm}
\begin{tabular}{l|c|c|c|c|c|c||c|c}
\hline
Method & \multicolumn{2}{c|}{Reddit} &  \multicolumn{2}{c|}{Wikipedia} & \multicolumn{2}{c||}{LastFM} & \multicolumn{2}{c}{Minimum \% improvement of \method over method} \\
& MRR & Recall@10 & MRR & Recall@10 & MRR & Recall@10 & MRR & Recall@10\\\hline
LSTM~\cite{zhu2017next} & 0.355 & 0.551 & 0.329 & 0.455 & 0.062 & 0.119 & 104.5\% & 54.6\% \\
Time-LSTM~\cite{zhu2017next} & 0.387 & 0.573 & 0.247 & 0.342 & 0.068 & 0.137 & 87.6\% & 48.7\% \\
RRN~\cite{wu2017recurrent} & \cellcolor{blue!10}0.603 & \cellcolor{blue!10}0.747 & \cellcolor{blue!10}0.522 &\cellcolor{blue!10} 0.617 & 0.089 & 0.182 & 20.4\% & 14.1\% \\
LatentCross~\cite{beutel2018latent} & 0.421 & 0.588 & 0.424 & 0.481 & \cellcolor{blue!10}0.148 & \cellcolor{blue!10}0.227 & 31.8\% & 35.2\%\\
CTDNE~\cite{nguyen2018continuous} & 0.165 & 0.257 & 0.035 & 0.056 & 0.01 & 0.01& 340.0\% & 231.5\% \\
DeepCoevolve~\cite{dai2016deep} & 0.171 & 0.275 & 0.515 & 0.563 & 0.019 & 0.039 & 44.8\% & 46.0\% \\
\method\ (proposed) &  \cellcolor{blue!25}\textbf{0.726} &  \cellcolor{blue!25}\textbf{0.852} &  \cellcolor{blue!25}\textbf{0.746} &  \cellcolor{blue!25}\textbf{0.822} &  \cellcolor{blue!25}\textbf{0.195} &  \cellcolor{blue!25}\textbf{0.307} & - & - \\\hline
\end{tabular}
\vspace{-2mm}
\end{table*}
}

\textbf{Baselines.}
We compare \method\ with six state-of-the-art algorithms spanning three algorithmic categories:
\begin{enumerate}
\item \textbf{Deep recurrent recommender models:} in this category, we compare with RRN~\cite{wu2017recurrent}, LatentCross~\cite{beutel2018latent}, Time-LSTM~\cite{zhu2017next}, and standard LSTM. These algorithms are state-of-the-art in recommender systems and generate dynamic user embeddings. We use Time-LSTM-3 cell for Time-LSTM as it performs the best in the original paper~\cite{zhu2017next}, and LSTM cells in RRN and LatentCross models. As is standard, we use the one-hot vector of items as inputs to these models.
\item \textbf{Dynamic co-evolution models:} here we compare with the state-of-the-art algorithm, DeepCoevolve~\cite{dai2016deep}, which has been shown to outperform other co-evolutionary point-process algorithms~\cite{trivedi2017know,wang2016coevolutionary}. We use 10 negative samples per interaction for computational tractability.
\item \textbf{Temporal network embedding models:} we compare \method\ with CTDNE~\cite{nguyen2018continuous} which is the state-of-the-art in generating embeddings from temporal networks. As it generates static embeddings, we generate new embeddings after each edge is added. We use uniform sampling of neighborhood as it performs the best in the original paper~\cite{nguyen2018continuous}.
\end{enumerate}

\vspace{-2mm}
\subsection{Experiment 1: Future interaction prediction}
\label{sec:exp1}
The prediction task here is: given all interactions till time $t$, which item will user $u$ interact with at time $t$ (out of all $|\mathcal{I}|$ items)?

We use three datasets in this experiments:\\
$\bullet$ \textbf{Reddit post dataset:} this public dataset consists of one month of posts made by users on subreddits~\cite{pushshift}. We selected the 1,000 most active subreddits as items and the 10,000 most active users. This results in 672,447 interactions. We convert the text of each post into a feature vector representing their LIWC categories~\cite{pennebaker2001linguistic}. \\
$\bullet$ \textbf{Wikipedia edits:} this public dataset is one month of edits made by edits on Wikipedia pages~\cite{wikidump}. We selected the 1,000 most edited pages as items and editors who made at least 5 edits as users (a total of 8,227 users). This generates 157,474 interactions. Similar to the Reddit dataset, we convert the edit text into a LIWC-feature vector. \\
$\bullet$ \textbf{LastFM song listens:} this public dataset has one month of who-listens-to-which song information~\cite{lastfm}. We selected all 1000 users and the 1000 most listened songs resulting in 1,293,103 interactions. In this dataset, interactions do not have features.

We select these datasets such that they vary in terms of users' repetitive behavior: in Wikipedia and Reddit, a user interacts with the same item consecutively in 79\% and 61\% interactions, respectively, while in LastFM, this happens in only 8.6\% interactions.

{
\small
\begin{table}
\caption{\label{tab:churn} \textbf{User state change prediction:} Table comparing the performance in terms of AUC of \method\ with state of the art algorithms. The {\color{blue!75}best algorithm} in each column is colored {\color{blue!75}blue} and the {\color{blue!20}second best is light blue}. \method\ outperforms the baselines by at least 12.63\% on average.}
\vspace{-2mm}
\begin{tabular}{l|c|c|c||c}
\hline
Method & Reddit & Wikipedia & MOOC & Mean improvement \\
& & & & of \method\ \\\hline
LSTM & 0.523 & 0.575 & 0.686 &23.08\% \\
Time-LSTM & 0.556 & 0.671 & \cellcolor{blue!10}0.711 & 12.63\% \\
RRN & \cellcolor{blue!10}0.586 & \cellcolor{blue!10}0.804 & 0.558 & 13.69\% \\
LatentCross & 0.574 & 0.628 & 0.686 & 15.62\% \\
DeepCoevolve & 0.577 & 0.663 & 0.671 & 13.94\% \\
\method & \cellcolor{blue!25}\textbf{0.599} &\cellcolor{blue!25} \textbf{0.831} & \cellcolor{blue!25}\textbf{0.756} & - \\\hline
\end{tabular}
\vspace{-3mm}
\end{table}
}
\textbf{Experimental setting.}
We use the first 80\% data to train, next 10\% to validate, and the final 10\% to test.
We measure the performance of the algorithms in terms of the mean reciprocal rank (MRR) and recall@10---MRR is the average of the reciprocal rank and recall@10 is the fraction of interactions in which the ground truth item is ranked in the top 10. Higher values for both are better.
For every interaction, the ranking of ground truth item is calculated with respect to all the items in the dataset.

For \method, items are ranked based on their $L_2$ distance from the predicted item embedding. The rank of the ground truth item is calculated in this ranked list.

\textbf{Results.}
Table~\ref{tab:interaction} compares the results of \method\ with the six state-of-the-art methods.
We observe that \method\ significantly outperforms all baselines in all datasets across both metrics on the three datasets.
Among the baselines, there is no clear winner---while RRN performs the better in Reddit and Wikipedia, LatentCross performs better in LastFM.
As CTDNE generates static embedding, its performance is low.
We calculate the percentage improvement of \method\ over the baseline as (performance of \method\ minus performance of baseline)/(performance of baseline). Across all datasets, the minimum improvement of \method\ is at least 20\% in terms of MRR and 14\% in terms of recall@10.
Please note that \method\ outperforms DeepCoevolve, the closest baseline in terms of the algorithm, by at least 44.8\% in MRR across all datasets.

Noticeably, we observe that \method\ performs well irrespective of how repetitive users are---the MRR at least 20.4\% higher in Wikipedia and Reddit (high repetition datasets), and at least 31.75\% higher in LastFM (low repetition dataset). This means \method\ is able to learn to balance personal preference with users' non-repetitive interaction behavior.

\vspace{-2mm}
\subsection{Experiment 2: User state change prediction}
\label{sec:exp2}
In this experiment, the task is to predict if an interaction will lead to a state change in user, particularly in two use cases: predicting if a user will be banned and predicting if a student will drop-out of a course.
Till a user is banned or drops-out, the label of the user is `0', and their last interaction has the label `1'. For users that are not banned or do not drop-out, the label is always `0'.
This is a highly challenging task as less than 1\% of the labels are `1`. 

\begin{figure}[t]
\centering
\includegraphics[width=0.7\columnwidth]{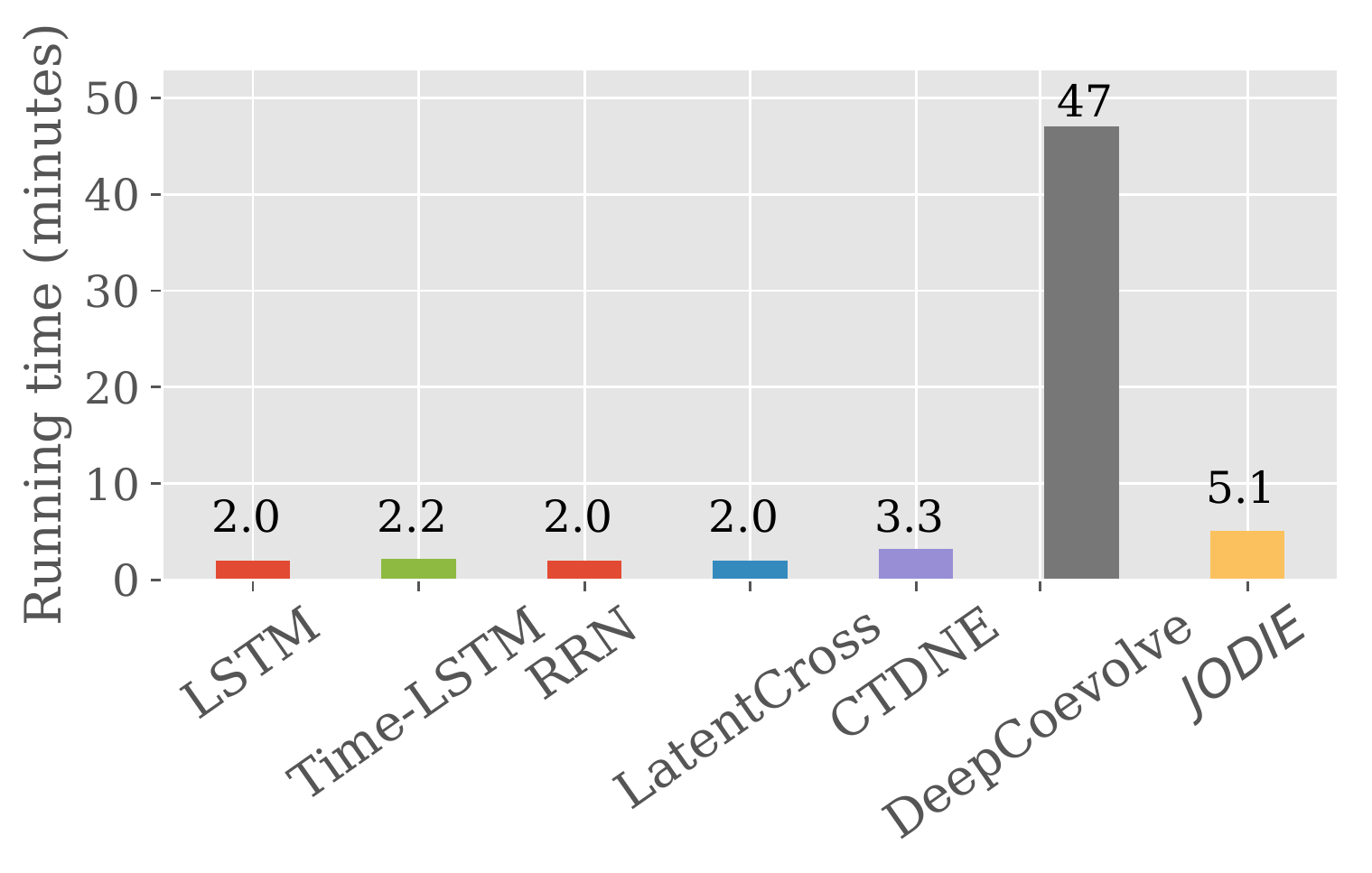}
\vspace{-4mm}
\caption{Figure compares the running time of \method\ and all baselines on the Reddit dataset. \method\ is 9.2$\times$ faster than DeepCoevolve and is comparable to the other baselines. \label{fig:runtime}}
\vspace{-3mm}
\end{figure}

\begin{figure*}[t]
\centering
\subfigure[\vspace{-3mm}]{
\includegraphics[width=0.22\textwidth]{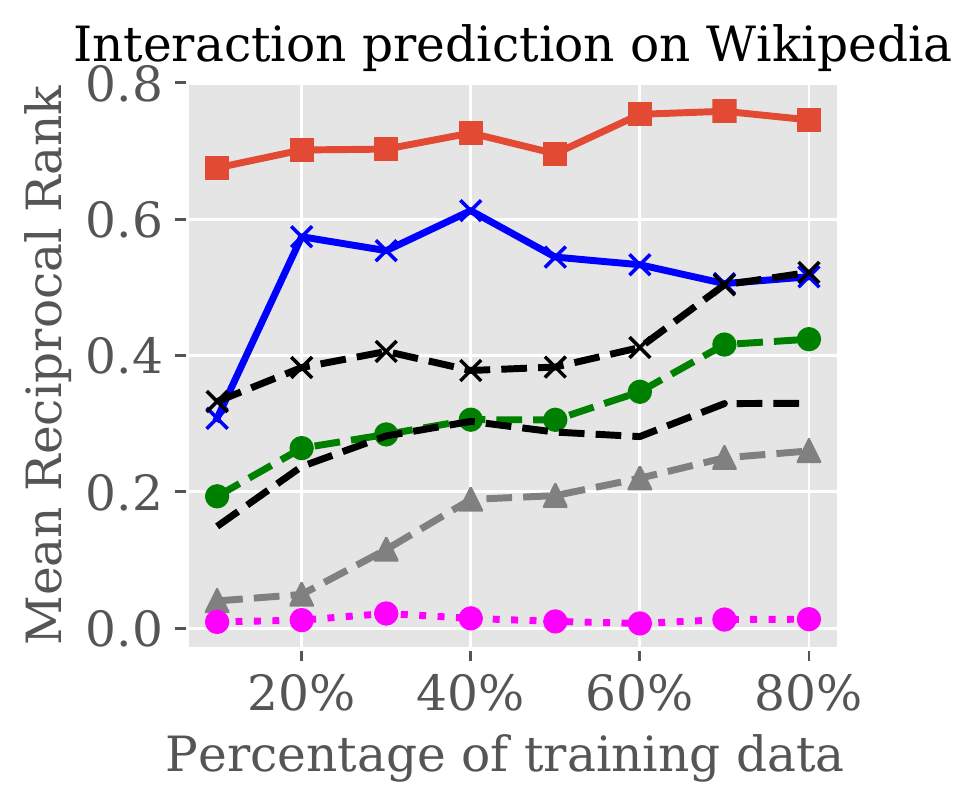}
}
\subfigure[\vspace{-3mm}]{
\includegraphics[width=0.21\textwidth]{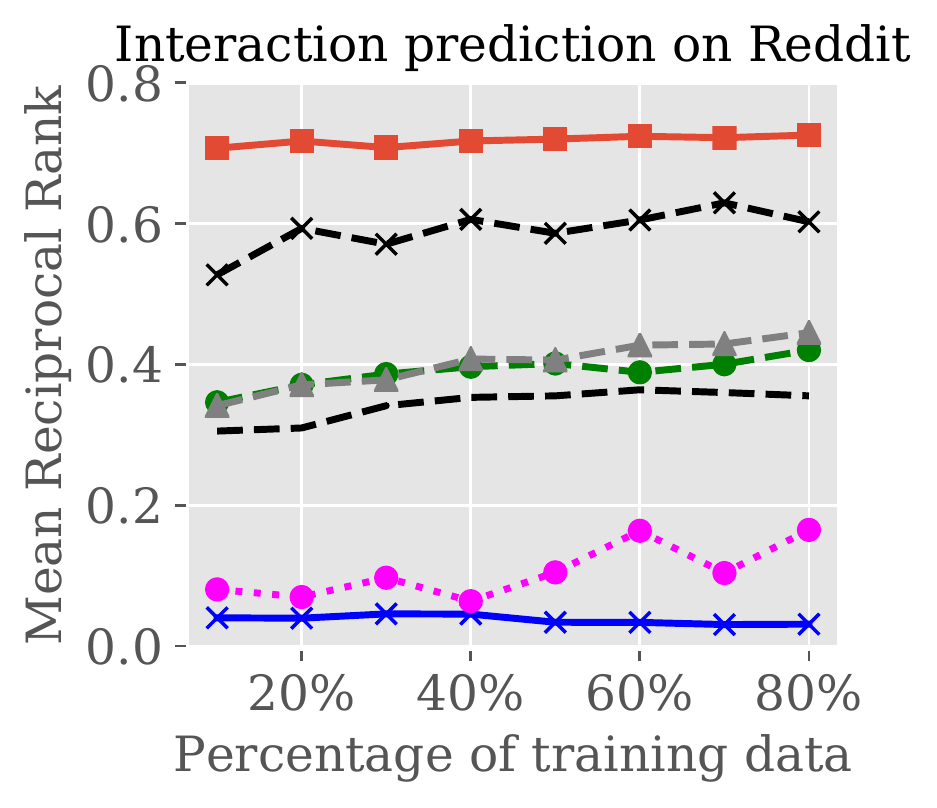}
}
\subfigure[\vspace{-3mm}]{
\includegraphics[width=0.21\textwidth]{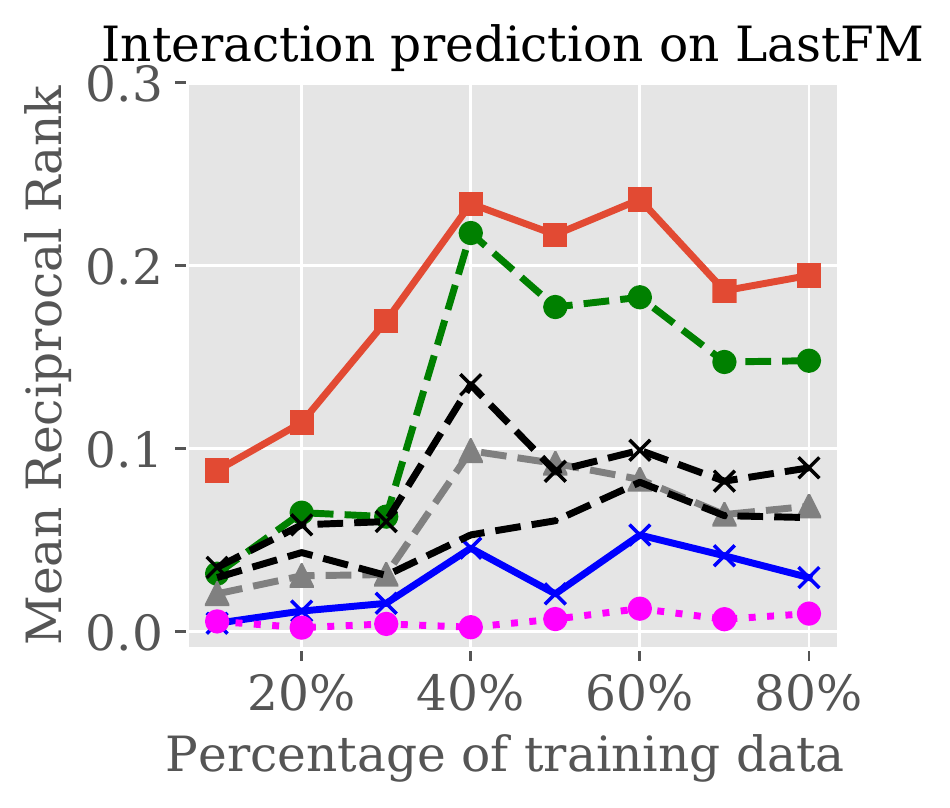}
}
\subfigure[\vspace{-3mm}]{
\includegraphics[width=0.21\textwidth]{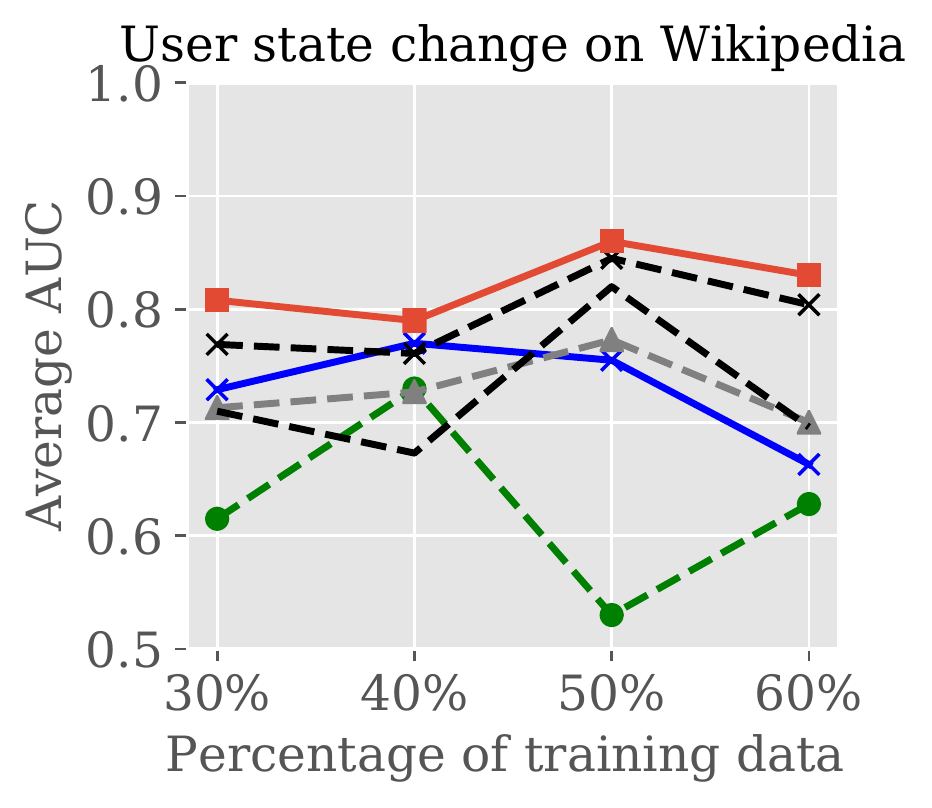}
}

\includegraphics[width=0.8\textwidth]{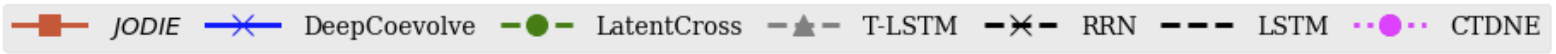}
\vspace{-2mm}
\caption{\textbf{Robustness of \method :} Figures (a--c) compare the mean reciprocal rank (MRR) of \method\ with baselines on interaction prediction task, by varying the training data size. Figure (d) shows the AUC of user state change prediction task by varying the training data size. We see \method\ consistently has the highest scores. \label{fig:interaction}}
\vspace{-4mm}
\end{figure*}

We use three datasets for this task: \\
$\bullet$ \textbf{Reddit bans:} Reddit post dataset (from Section~\ref{sec:exp1}) with ground-truth labels of banned users from Reddit
This gives 366 true labels among 672,447 interactions (= 0.05\%).\\
$\bullet$ \textbf{Wikipedia bans:} Wikipedia edit data (from Section~\ref{sec:exp1}) with public ground-truth labels of banned users~\cite{wikidump}. This results in 217 positive labels among 157,474 interactions (= 0.14\%).  \\
$\bullet$ \textbf{MOOC student drop-out:} this public dataset consists of actions, e.g., viewing a video, submitting an answer, etc., done by students on a MOOC online course~\cite{kddcup}. This dataset consists of 7,047 users interacting with 98 items (videos, answers, etc.) resulting in over 411,749 interactions. There are 4,066 drop-out events (= 0.98\%).

\textbf{Experimental setting.}
In this experiment, we train the models on the first 60\% interactions, validate on the next 20\%, and test on the last 20\% interactions.
We evaluate the models using the area under the curve metric (AUC), a standard metric in the tasks with highly imbalanced labels.

For the baselines, we train a logistic regression classifier on the training data using the dynamic user embedding as input.
As always, for all models, we report the test AUC for the epoch with the highest validation AUC.

\textbf{Results.}
Table~\ref{tab:churn} compares the performance of \method\ on the three datasets with the baseline models.
We see that \method\ outperforms the baselines by at least 12\% on average in predicting user state change across all datasets.
\method\ outperforms RRN, the closest competitor in the ban prediction task, by at least 2.2\% while it outperforms RRN by 28\% in the student drop-out task.
Note that DeepCoevolve, which is the most similar baseline algorithmically, is outperformed by 13.9\% by \method\ on average.
Thus, \method\ consistently performs the best across various datasets.

\vspace{-2mm}
\subsection{Experiment 3: Runtime experiment}
\label{sec:exp-tbatch}
\vspace{-1mm}
Here we compare the running time of \method\ with the baseline algorithms.
Algorithmically, the DeepCoevolve is the closest to \method\ as it also trains two mutually-recursive RNNs. The other methods train only one RNN and are therefore easily scalable.

Figure~\ref{fig:runtime} shows the running time (in minutes) of one epoch of the Reddit dataset.\footnote{We ran the experiment on one NVIDIA Titan X Pascal GPUs with 12Gb of RAM at 10Gbps speed.}
We find that \method\ is 9.2$\times$ faster than DeepCoevolve (its closest algorithmic competitor). At the same time, the running time of \method\ is comparable to the other baselines that only use one RNN in their model. This shows that \method\ is able to train the mutually-recursive model in equivalent time as non-mutually-recursive models, because of the use of the \batching\ training batching algorithm.

In addition, we find that \method\ without \batching\ took 43.53 minutes while \method\ with \batching\ took 5.13 minutes. Thus, \batching\ results in 8.4$\times$ speed-up.

\vspace{-2mm}
\subsection{Experiment 4: Robustness to the proportion of training data}
\label{sec:exp-training}
\vspace{-1mm}
In this experiment, we validate the robustness of \method\ by varying the percentage of training data and comparing the performance of the algorithms in both the tasks of future interaction prediction and user state change prediction.

For the next item prediction, we vary the training data percentage from 10\% to 80\%. In each case, we take the 10\% interactions after the training data as validation and the next 10\% interactions next as testing. This is done to compare the performance on the same testing data size.
Figures~\ref{fig:interaction}(a--c) show the change in mean reciprocal rank (MRR) of all the algorithms on the three datasets, as the training data size is increased.
We note that the performance of \method\ is stable and does not vary much across the data points.
Moreover, \method\ consistently outperforms the baseline models by a significant margin (by a maximum of 33.1\%).

We make similar observations in user state change prediction task. Here, we vary training data percents to 20\%, 40\%, and 60\%, and in each case take the following 20\% interactions as validation and the next 20\% interactions as the test.
Figure~\ref{fig:interaction}(d) shows the AUC of all the algorithms on the Wikipedia dataset. Other datasets have similar results.
Again, we find that \method\ is stable and consistently outperforms the baselines, irrespective of the training data size.

\begin{figure}[t]
\centering
\includegraphics[width=0.5\columnwidth]{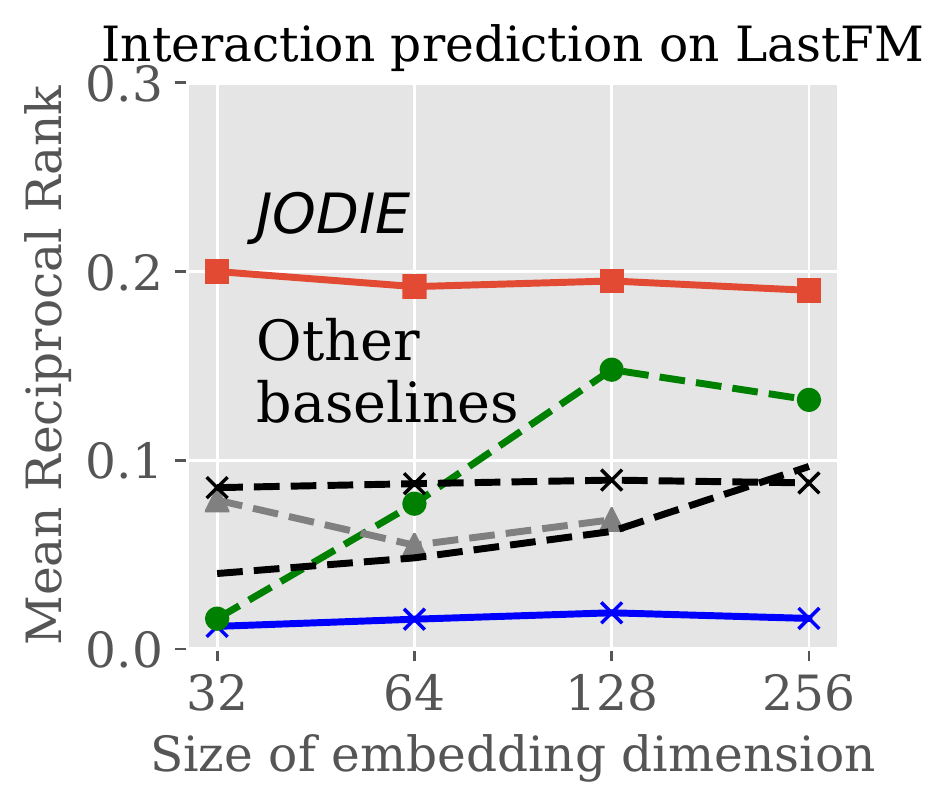}
\vspace{-4mm}
\caption{\textbf{Robustness to dynamic embedding size:} The performance of \method\ is stable with the change in dynamic embedding size, for the task of interaction prediction on LastFM dataset. Please refer to the legend in Figure 5. \label{fig:embed}
}
\vspace{-2mm}
\end{figure}

\vspace{-2mm}
\subsection{Experiment 5: Embedding size}
\label{sec:exp-embsize}
Finally, we validate the effect of the dynamic embedding size on the predictions.
To do this, we vary the dynamic embedding dimension from 32 to 256 and calculate the mean reciprocal rank for interaction prediction on the LastFM dataset. The effect on other datasets is similar. 
The resulting figure is showing in Figure~\ref{fig:embed}.
We find that the embedding dimension size has little effect on the performance of \method\ and it performs the best overall.
Interestingly, improvement in \method\ is higher for smaller embedding dimensions.
This is because \method uses both the static and the dynamic embedding for prediction, which gives it the power to learn from both parts.

\section{Conclusions}
In this paper, we proposed a coupled recurrent neural network model called \method\ that learns dynamic embeddings of users and items from a sequence of temporal interactions.
\method\ learns to predict the future embeddings of users and items, which leads it to give better prediction performance of future user-item interactions and change in user state. We also presented a training data batching method that makes \method\ an order of magnitude faster than similar baselines.

There are several directions for future work. Learning embeddings for individual users and items is expensive, and one could learn trajectories for groups of users or items to reduce the number of parameters. Another direction is characterizing the trajectories to cluster similar entities. Finally, an innovative direction would be to design new items based on missing predicted items that many users are likely to interact with.

\xhdr{Acknowledgements}
JL is a Chan Zuckerberg Biohub investigator.
This research has been supported in part by
NSF OAC-1835598, DARPA MCS, DARPA ASED, ARO MURI,
Amazon, Boeing, Docomo, Hitachi, JD, Siemens, and Stanford Data Science Initiative.
We thank Sagar Honnungar for help with the initial phase of the project.

\bibliographystyle{abbrv}
\balance
\bibliography{bib/refs}

\begin{thebibliography}{10}

\bibitem{kddcup}
Kdd cup 2015.
\newblock \url{https://biendata.com/competition/kddcup2015/data/}.

\bibitem{pushshift}
Reddit data dump.
\newblock \url{http://files.pushshift.io/reddit/}.

\bibitem{wikidump}
Wikipedia edit history dump.
\newblock \url{https://meta.wikimedia.org/wiki/Data_dumps}.

\bibitem{agrawal2014big}
D.~Agrawal, C.~Budak, A.~El~Abbadi, T.~Georgiou, and X.~Yan.
\newblock Big data in online social networks: user interaction analysis to
  model user behavior in social networks.
\newblock In {\em DNIS}, 2014.

\bibitem{DBLP:conf/asunam/ArnouxTL17}
T.~Arnoux, L.~Tabourier, and M.~Latapy.
\newblock Combining structural and dynamic information to predict activity in
  link streams.
\newblock In {\em ASONAM}, 2017.

\bibitem{DBLP:journals/corr/abs-1804-01465}
T.~Arnoux, L.~Tabourier, and M.~Latapy.
\newblock Predicting interactions between individuals with structural and
  dynamical information.
\newblock {\em CoRR}, 2018.

\bibitem{baytas2017patient}
I.~M. Baytas, C.~Xiao, X.~Zhang, F.~Wang, A.~K. Jain, and J.~Zhou.
\newblock Patient subtyping via time-aware lstm networks.
\newblock In {\em KDD}, 2017.

\bibitem{beutel2018latent}
A.~Beutel, P.~Covington, S.~Jain, C.~Xu, J.~Li, V.~Gatto, and E.~H. Chi.
\newblock Latent cross: Making use of context in recurrent recommender systems.
\newblock In {\em WSDM}, 2018.

\bibitem{cheng2017anyone}
J.~Cheng, M.~Bernstein, C.~Danescu-Niculescu-Mizil, and J.~Leskovec.
\newblock Anyone can become a troll: Causes of trolling behavior in online
  discussions.
\newblock In {\em CSCW}, 2017.

\bibitem{cheng2017predicting}
J.~Cheng, C.~Lo, and J.~Leskovec.
\newblock Predicting intent using activity logs: How goal specificity and
  temporal range affect user behavior.
\newblock In {\em WWW}, 2017.

\bibitem{dai2016deep}
H.~Dai, Y.~Wang, R.~Trivedi, and L.~Song.
\newblock Deep coevolutionary network: Embedding user and item features for
  recommendation.
\newblock {\em arXiv:1609.03675}, 2016.

\bibitem{DBLP:conf/kdd/DuDTUGS16}
N.~Du, H.~Dai, R.~Trivedi, U.~Upadhyay, M.~Gomez{-}Rodriguez, and L.~Song.
\newblock Recurrent marked temporal point processes: Embedding event history to
  vector.
\newblock In {\em KDD}, 2016.

\bibitem{DBLP:conf/nips/FarajtabarWGLZS15}
M.~Farajtabar, Y.~Wang, M.~Gomez{-}Rodriguez, S.~Li, H.~Zha, and L.~Song.
\newblock {COEVOLVE:} {A} joint point process model for information diffusion
  and network co-evolution.
\newblock In {\em NeurIPS}, 2015.

\bibitem{DBLP:journals/kbs/GoyalF18}
P.~Goyal and E.~Ferrara.
\newblock Graph embedding techniques, applications, and performance: {A}
  survey.
\newblock {\em Knowledge Based Systems}, 151:78--94, 2018.

\bibitem{goyal2018dyngem}
P.~Goyal, N.~Kamra, X.~He, and Y.~Liu.
\newblock Dyngem: Deep embedding method for dynamic graphs.
\newblock {\em arXiv:1805.11273}, 2018.

\bibitem{grover2016node2vec}
A.~Grover and J.~Leskovec.
\newblock node2vec: Scalable feature learning for networks.
\newblock In {\em KDD}, 2016.

\bibitem{DBLP:journals/debu/HamiltonYL17}
W.~L. Hamilton, R.~Ying, and J.~Leskovec.
\newblock Representation learning on graphs: Methods and applications.
\newblock {\em {IEEE} Data Engineering Bulletin}, 40(3):52--74, 2017.

\bibitem{lastfm}
B.~Hidasi and D.~Tikk.
\newblock Fast als-based tensor factorization for context-aware recommendation
  from implicit feedback.
\newblock In {\em ECML}, 2012.

\bibitem{iba2010analyzing}
T.~Iba, K.~Nemoto, B.~Peters, and P.~A. Gloor.
\newblock Analyzing the creative editing behavior of wikipedia editors: Through
  dynamic social network analysis.
\newblock {\em Procedia-Social and Behavioral Sciences}, 2(4):6441--6456, 2010.

\bibitem{julier1997new}
S.~J. Julier and J.~K. Uhlmann.
\newblock New extension of the kalman filter to nonlinear systems.
\newblock In {\em Signal processing, sensor fusion, and target recognition VI},
  volume 3068, pages 182--194, 1997.

\bibitem{DBLP:journals/corr/abs-1711-10967}
R.~R. Junuthula, M.~Haghdan, K.~S. Xu, and V.~K. Devabhaktuni.
\newblock The block point process model for continuous-time event-based dynamic
  networks.
\newblock {\em CoRR}, 2017.

\bibitem{DBLP:conf/icwsm/Junuthula0D18}
R.~R. Junuthula, K.~S. Xu, and V.~K. Devabhaktuni.
\newblock Leveraging friendship networks for dynamic link prediction in social
  interaction networks.
\newblock In {\em ICWSM}, 2018.

\bibitem{kloft2014predicting}
M.~Kloft, F.~Stiehler, Z.~Zheng, and N.~Pinkwart.
\newblock Predicting mooc dropout over weeks using machine learning methods.
\newblock In {\em EMNLP}, 2014.

\bibitem{kumar2018community}
S.~Kumar, W.~L. Hamilton, J.~Leskovec, and D.~Jurafsky.
\newblock Community interaction and conflict on the web.
\newblock In {\em The World Wide Web Conference}, 2018.

\bibitem{kumar2018rev2}
S.~Kumar, B.~Hooi, D.~Makhija, M.~Kumar, C.~Faloutsos, and V.~Subrahmanian.
\newblock Rev2: Fraudulent user prediction in rating platforms.
\newblock In {\em WSDM}, 2018.

\bibitem{kumar2015vews}
S.~Kumar, F.~Spezzano, and V.~Subrahmanian.
\newblock Vews: A wikipedia vandal early warning system.
\newblock In {\em KDD}, 2015.

\bibitem{leskovec2014mining}
J.~Leskovec, A.~Rajaraman, and J.~D. Ullman.
\newblock {\em Mining of massive datasets}.
\newblock Cambridge university press, 2014.

\bibitem{DBLP:conf/cikm/LiDHTCL17}
J.~Li, H.~Dani, X.~Hu, J.~Tang, Y.~Chang, and H.~Liu.
\newblock Attributed network embedding for learning in a dynamic environment.
\newblock In {\em CIKM}, 2017.

\bibitem{DBLP:journals/access/LiZYZY18}
T.~Li, J.~Zhang, P.~S. Yu, Y.~Zhang, and Y.~Yan.
\newblock Deep dynamic network embedding for link prediction.
\newblock {\em {IEEE} Access}, 6:29219--29230, 2018.

\bibitem{DBLP:conf/sdm/LiDLLGZ14}
X.~Li, N.~Du, H.~Li, K.~Li, J.~Gao, and A.~Zhang.
\newblock A deep learning approach to link prediction in dynamic networks.
\newblock In {\em SDM}, 2014.

\bibitem{liyanagunawardena2013moocs}
T.~R. Liyanagunawardena, A.~A. Adams, and S.~A. Williams.
\newblock Moocs: A systematic study of the published literature 2008-2012.
\newblock {\em The International Review of Research in Open and Distributed
  Learning}, 14(3):202--227, 2013.

\bibitem{DBLP:journals/corr/abs-1810-10627}
Y.~Ma, Z.~Guo, Z.~Ren, Y.~E. Zhao, J.~Tang, and D.~Yin.
\newblock Dynamic graph neural networks.
\newblock {\em CoRR}, abs/1810.10627, 2018.

\bibitem{nguyen2018continuous}
G.~H. Nguyen, J.~B. Lee, R.~A. Rossi, N.~K. Ahmed, E.~Koh, and S.~Kim.
\newblock Continuous-time dynamic network embeddings.
\newblock In {\em WWW BigNet workshop}, 2018.

\bibitem{DBLP:conf/recsys/PalovicsBKKF14}
R.~P{\'{a}}lovics, A.~A. Bencz{\'{u}}r, L.~Kocsis, T.~Kiss, and E.~Frig{\'{o}}.
\newblock Exploiting temporal influence in online recommendation.
\newblock In {\em RecSys}, 2014.

\bibitem{pennebaker2001linguistic}
J.~W. Pennebaker, M.~E. Francis, and R.~J. Booth.
\newblock Linguistic inquiry and word count: Liwc 2001.
\newblock {\em Mahway: Lawrence Erlbaum Associates}, 71(2001):2001, 2001.

\bibitem{DBLP:conf/wsdm/QiuDMLWT18}
J.~Qiu, Y.~Dong, H.~Ma, J.~Li, K.~Wang, and J.~Tang.
\newblock Network embedding as matrix factorization: Unifying deepwalk, line,
  pte, and node2vec.
\newblock In {\em WSDM}, 2018.

\bibitem{raghavan2014modeling}
V.~Raghavan, G.~Ver~Steeg, A.~Galstyan, and A.~G. Tartakovsky.
\newblock Modeling temporal activity patterns in dynamic social networks.
\newblock {\em IEEE TCSS}, 1(1):89--107, 2014.

\bibitem{DBLP:journals/corr/abs-1804-05755}
M.~Rahman, T.~K. Saha, M.~A. Hasan, K.~S. Xu, and C.~K. Reddy.
\newblock Dylink2vec: Effective feature representation for link prediction in
  dynamic networks.
\newblock {\em CoRR}, 2018.

\bibitem{DBLP:journals/corr/abs-1710-00818}
S.~Sajadmanesh, J.~Zhang, and H.~R. Rabiee.
\newblock Continuous-time relationship prediction in dynamic heterogeneous
  information networks.
\newblock {\em CoRR}, 2017.

\bibitem{DBLP:conf/cosn/SedhainSXKTC13}
S.~Sedhain, S.~Sanner, L.~Xie, R.~Kidd, K.~Tran, and P.~Christen.
\newblock Social affinity filtering: recommendation through fine-grained
  analysis of user interactions and activities.
\newblock In {\em COSN}, 2013.

\bibitem{trivedi2017know}
R.~Trivedi, H.~Dai, Y.~Wang, and L.~Song.
\newblock Know-evolve: Deep temporal reasoning for dynamic knowledge graphs.
\newblock In {\em ICML}, 2017.

\bibitem{trivedi2018representation}
R.~Trivedi, M.~Farajtbar, P.~Biswal, and H.~Zha.
\newblock Representation learning over dynamic graphs.
\newblock {\em arXiv:1803.04051}, 2018.

\bibitem{walker2015complex}
P.~B. Walker, S.~G. Fooshee, and I.~Davidson.
\newblock Complex interactions in social and event network analysis.
\newblock In {\em SBP-BRiMS}, 2015.

\bibitem{wang2016coevolutionary}
Y.~Wang, N.~Du, R.~Trivedi, and L.~Song.
\newblock Coevolutionary latent feature processes for continuous-time user-item
  interactions.
\newblock In {\em NeurIPS}, 2016.

\bibitem{wu2017recurrent}
C.-Y. Wu, A.~Ahmed, A.~Beutel, A.~J. Smola, and H.~Jing.
\newblock Recurrent recommender networks.
\newblock In {\em WSDM}, 2017.

\bibitem{yang2013turn}
D.~Yang, T.~Sinha, D.~Adamson, and C.~P. Ros{\'e}.
\newblock Turn on, tune in, drop out: Anticipating student dropouts in massive
  open online courses.
\newblock In {\em NeurIPS Data-driven education workshop}, 2013.

\bibitem{you2019hierarchical}
J.~You, Y.~Wang, A.~Pal, P.~Eksombatchai, C.~Rosenburg, and J.~Leskovec.
\newblock Hierarchical temporal convolutional networks for dynamic recommender
  systems.
\newblock In {\em The World Wide Web Conference}, 2019.

\bibitem{zhang2017deep}
S.~Zhang, L.~Yao, and A.~Sun.
\newblock Deep learning based recommender system: A survey and new
  perspectives.
\newblock {\em arXiv:1707.07435}, 2017.

\bibitem{zhang2017learning}
Y.~Zhang, Y.~Xiong, X.~Kong, and Y.~Zhu.
\newblock Learning node embeddings in interaction graphs.
\newblock In {\em CIKM}, 2017.

\bibitem{zhou2018dynamic}
L.-k. Zhou, Y.~Yang, X.~Ren, F.~Wu, and Y.~Zhuang.
\newblock Dynamic network embedding by modeling triadic closure process.
\newblock In {\em AAAI}, 2018.

\bibitem{zhu2016scalable}
L.~Zhu, D.~Guo, J.~Yin, G.~Ver~Steeg, and A.~Galstyan.
\newblock Scalable temporal latent space inference for link prediction in
  dynamic social networks.
\newblock {\em IEEE TKDE}, 28(10):2765--2777, 2016.

\bibitem{zhu2017next}
Y.~Zhu, H.~Li, Y.~Liao, B.~Wang, Z.~Guan, H.~Liu, and D.~Cai.
\newblock What to do next: modeling user behaviors by time-lstm.
\newblock In {\em IJCAI}, 2017.

\end{thebibliography}

\newpage
\appendix

\section{Appendix}
Here we describe some technical details of the model.

The code and datasets are available on the project website: \\\texttt{https://snap.stanford.edu/jodie}.

We coded all the models and the baselines in PyTorch. Table~\ref{tab:app2} mentions the dataset details and Table~\ref{tab:app1} mentions the model parameters.

\begin{table}
\caption{Table with model parameters. \label{tab:app1}}
\begin{tabular}{|c|c|}
\hline
Parameter & Value \\\hline
Optimizer & Adam\\
Learning rate & 1e-3 \\
Model weight decay & 1e-5 \\
Dynamic embedding size & 128 \\
Number of epochs & 50 \\\hline
\multicolumn{2}{|c|}{Future interaction prediction experiment}\\\hline
Training data percent & 80\% \\
Validation data percent & 10\% \\
Test data percent & 10\% \\\hline
\multicolumn{2}{|c|}{User state change  experiment}\\\hline
Training data percent & 60\% \\
Validation data percent & 20\% \\
Test data percent & 20\% \\\hline
\end{tabular}
\end{table}

\begin{table}
\caption{Table with dataset information. \label{tab:app2}}
\begin{tabular}{|c|c|c|c|c|c|}
\hline
Data & Users & Items & Interactions & State & Action \\
& & & & Changes & Repetition\\\hline
Reddit & 10,000 & 984 & 672,447 & 366 & 79\% \\
Wikipedia & 8,227 & 1,000 & 157,474 &  217 & 61\%\\
LastFM & 980 & 1,000 & 1,293,103 & - & 8.6\% \\
MOOC & 7,047 & 97 & 411,749 & 4,066 & -\\\hline
\end{tabular}
\end{table}

\end{document}